\documentclass[aps,pra,twocolumn,floats]{revtex4}
\usepackage{graphicx}
\usepackage{amsmath}

\usepackage{epsfig}
\usepackage{amssymb}

\begin{document}

\title{Control of catalytic activity of proteins {\it in vivo} 
by nanotube ropes \\ excited with infrared light}

\author{Petr Kr\'al}

\address{Department of Chemical Physics, Weizmann Institute of Science,
         76100 Rehovot, Israel}

\begin{abstract}
We discuss the possibility of controlling biological systems, by exciting 
in the near infrared region {\it hybrid} metallic nanotube ropes, dressed 
with proteins and embedded in the biosystems. If one nanotube, in a double-tube 
rope, is filled with metallofullerenes and the other is empty, the two tubes
change their opposite equilibrium charging during the irradiation. The 
resulting change of the local electric field can deform proteins attached 
to the tubes, and change their catalytic properties. 
\end{abstract}

\maketitle


\section{Introduction}

Over billions of years, a fascinating internal and external complexity 
evolved in myriads of competing biological species \cite{Kral}.  For example, 
many bacteria and multi-cellular organisms use structure-sensitive proteins 
to biomineralize nanocrystals \cite{Bauerlein}, that form unique nanodevices. 
These {\it cold}-growth techniques \cite{growth} could greatly complement 
growth methods developed by humans. In general, biosystems and artificial 
nanosystems can coexist and supplement each other in a number of other 
directions, in particular, during the release of drugs \cite{Uhrich}. 
Their coevolution can lead to the formation of new hybrid {\it bio-nano} 
(BIONA) systems, with unprecedented organizational and functional level. 

In this Letter, we discuss possible forms of communication between BIONA 
components. The (direct) ``talk" from the nanosystem to the 
biosystem could be realized by controlling catalytic activity of its 
proteins. The (backward) ``talk" could be done by the cells, if they 
change their local micro-environment or emit (electromagnetic) 
signals. The direct talk, that we explore here in more details,
could be based on electrochemical 
methods, used in bioelectronics \cite{Willner}. Unfortunately, these 
techniques require the presence of electrodes \cite{Mrksich}, that 
cannot be easily applied inside cells. More promising is thus their 
combination with {\it contactless} methods. Optical techniques \cite{PUMA} 
are convenient for their selectivity, but the sensitive interior of cells
prohibits the use of large optical frequencies that can easily manipulate 
chemical bonds.  We could thus use the fact that cells are transparent
(up to 5 mm thick samples) for the {\it near-infrared} (NIR) radiation
(0.74-1.2$\mu$m).

The activity of biosystems could be manipulated via artificial nanosystems, 
embedded in them, that absorb in the NIR region. This approach is followed 
in photodynamic therapies, where tumor cells are destructed chemically, via 
NIR excitation of porphyrin-based molecules with many extended electronic 
states \cite{MacDonald}. Similar results can be obtained by heating the system 
locally with ultrasound or microwave radiation or via NIR-radiation heated 
metallic 
nanoparticles \cite{Pitsillides}. Silver and gold nanoparticles \cite{Sun}, 
that can be produced biologically \cite{Klaus}, are naturally excellent 
candidates for use in bio-control. In order to control tinny cellular sections,
one could also think of using nanotubes. Metallic C nanotubes form sensitive 
detectors in liquid environments \cite{DRAG}, and when dressed with 
bio-molecules, via structure-selective \cite{Attached11,Attached12} or 
less specific hydrophobic coupling \cite{Attached21}, they 
can work as sensitive biosensors \cite{Attached22}.

\section{NIR-radiation control of protein activity}

The bio-control could be elegantly realized by a {\it hybrid} nanotube rope,
heated by NIR-radiation, that is formed of two adjacent metallic C nanotubes, 
where one is (peapod) 
filled with {\it metallofullerenes} \cite{Luzzi} and the other is empty.  
In peapods, electrons can be transferred to/from the C$_{60}$ fullerenes under 
electric bias \cite{Kavan}. In isolated metallofullerenes like Dy@C$_{82}$, 
several electrons are passed from Dy to C$_{82}$. When these are used in 
a peapod, the transferred electrons can be passed further to the nanotube
\cite{Roth}. In a double-rope formed by this peapod and a ``twin" (empty) 
nanotube, the last would absorb the excessive charge too, so the 
two would become oppositely charged. This process can be partly 
{\it inverted} at elevated temperatures \cite{Roth}, since the 
metallofullerenes have levels close to the Fermi level \cite{DLee,Cho}. 

\begin{figure}
\vspace{12mm}
\hspace*{0mm} 
{\hbox{\epsfxsize=72mm \epsffile{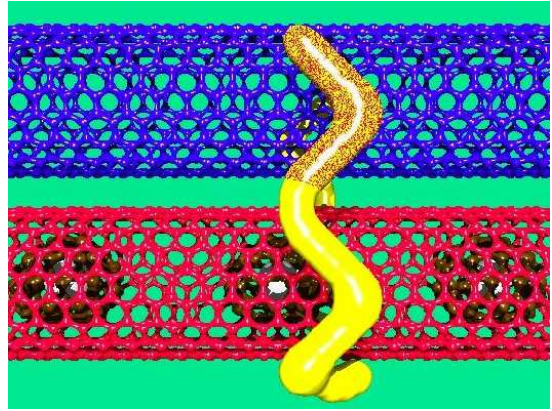} }
\vspace{-0mm} }
\caption{Scheme of a hybrid nanotube system that controls protein activity 
{\it in vivo} by near-infrared radiation, as described in the text. }
\label{BIONA1}
\end{figure}

In Fig.~\ref{BIONA1}, we schematically show the NIR-control of protein
(enzyme) activity, based on this process. The NIR excitation heats the
two nanotubes, so that electrons, released in equilibrium from the fullerenes
to the peapod \cite{Roth} and the twin tube, become transferred {\it back}.
This transfer is accompanied by recharging of the tubes, and the resulting 
change of the local electric field causes deformation of proteins 
\cite{Gerstein,Field}, that are selectively attached to the nanotubes. 
Their new conformation can have a largely different catalytic activity 
\cite{WillnerB}. The system thus works in an opposite way than some 
biosensors \cite{Benson}, where antibodies bind to proteins attached to 
material surfaces, bend them, and thus change the surface electric parameters. 
We can tune the system by using different nanotubes, fullerenes and their 
filling, and especially proteins, to be controlled. The attached proteins,
that in general could be much bigger than the tubular system, might help 
to dissolve the hydrophobic nanotubes in water.

\section{Modeling of the control}

We consider that the system is formed by two metallic (10,10) carbon nanotubes 
of the radius $r_T\approx 0.68$ nm, where {\it one} of them is the peapod. In a 
double-rope \cite{Stahl,Mele}, their centers are separated by $D_T\approx 1.7$ 
nm, which determines the tunneling time, $\tau_t \approx 1$ ps, of electrons 
between the tubes. The fullerenes are separated one from another by 
$d_F\approx 2$ nm \cite{Cho}, and their charging strongly but locally 
deforms electronic bands of the peapod \cite{DLee}. 

We can excite the metallic nanotubes at deliberate NIR frequencies.
Their absorbtion fits the Drude formula for the dielectric function 
\cite{Lee}
\begin{eqnarray}
\varepsilon(\omega)=\varepsilon_{\infty}
\Bigl(1-\frac{\omega_p^2}{\omega^2+i\omega/\tau} \Bigr)\, ,
\label{epsilon}
\end{eqnarray}
where $\hbar\omega_p=0.86$ eV is the plasma frequency, $\varepsilon_{\infty}
=4.6$ and $\tau=5\times 10^{-15}$ s. The irradiation, needed to induce
reabsorbtion of electrons by the fullerenes, would have to heat the nanotubes 
by several tens of degrees \cite{Roth}. This should not be harmful 
{\it in vivo}, if we use, for example, short (few nanometers long) 
nanotube {\it capsules} \cite{Tomanek}, and isolate them in liposomes
\cite{Haran}, that can be delivered to the cells by special techniques.

In the lack of available data, we assume here that one electron {\it per}
fullerene is reabsorbed during the irradiation, and $\approx 20\, \%$ of 
those come from the empty tube.  This gives the NIR-radiation induced 
{\it recharging} 
density $\sigma=0.2\,e/d_F\approx 0.1\, e$ nm$^{-1}$. From this $\sigma$, we 
can calculate the change of electric field between the two tubes. If we assume 
that these two are {\it ideal} metallic cylinders of length L, their electric 
capacity is \cite{Slater} $(\varepsilon =\varepsilon_0\, \varepsilon_r)$ 
\begin{eqnarray}
C=\frac{\pi\varepsilon L}{{\rm cosh}^{-1}(D_T/2\, r_T)}\, . \ \ \
\label{cap}
\end{eqnarray}
Thus, the potential difference between the nanotubes due to the induced 
charge transfer is 
\begin{eqnarray}
\Delta \varphi=\sigma L/C\approx 0.1\ {\rm V} \, ,
\label{pot}
\end{eqnarray}
where we use the permitivity of water $\varepsilon_r \approx 4.6$. Similar 
voltage was used, for example, in manipulation of proteins attached to metals 
\cite{WillnerB}. 

\begin{figure}
\vspace{-45mm}
\hspace*{-3mm} 
{\hbox{\epsfxsize=105mm \epsffile{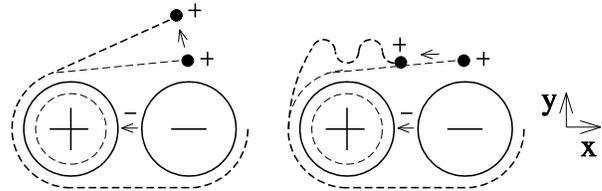} }
\vspace{-6mm} }
\caption{Scheme of two (hinge and shear) configurations for the protein 
control, as described in the text. }
\label{BIONA2}
\end{figure}

Activation of the attached protein by this NIR-radiation induced potential can 
be realized by moving a {\it charged tip} of one of its domains \cite{Benson} 
(see Fig.~\ref{BIONA2}). We model this process, by evaluating first the 
potential energy of a charge $q$ at the position ${\bf r}=(x,y)$. The two 
cylinders with charge densities $\sigma$ and $-\sigma$ have their centers 
are at ${\bf r}_1$ and ${\bf r}_2$, respectively. The potential energy of 
the charge $q$ is formed by the {\it direct} Coulombic component \cite{Slater} 
\begin{eqnarray}
V_C({\bf r})= -\frac{\sigma \, q\, }{2\pi\varepsilon\, } \, 
\ln\left(\frac{|{\bf r}-{\bf r}_1|}{|{\bf r}-{\bf r}_2|} \right)\, ,
\label{VC}
\end{eqnarray}
that can be either positive or negative, depending on which of the oppositely
charged tubes is closer. It also has a negative {\it screening} component 
\cite{TIS}, originating in the reflection of the external charge in the 
metallic tube, that close to the surface of both tubes has the form 
\begin{eqnarray}
V_S({\bf r}) \approx -\frac{q^2}{16\pi\varepsilon}\left(
\frac{1}{|{\bf r}-{\bf r}_1|-r_T}+
\frac{1}{|{\bf r}-{\bf r}_2|-r_T}\right)\, . 
\label{VS}
\end{eqnarray}
Here, we simply add the screening potentials of the two tubes, neglecting 
thus multiple reflections.

Typically, structural domains in proteins perform {\it hinge} or {\it shear} 
motion \cite{Gerstein,Field}. These domains are often formed by (rigid) 
$\alpha$-helices, connected by (flexible) $\beta$-sheets. 
The structures (conformations) of deformed proteins in nature are usually 
close in energies, so that they can be flipped over by room temperature 
energies k$_{B}$T \cite{Gerstein}. The conformations of the externally 
controlled proteins should have different catalytic properties and
be more energetically distant, so that they are not changed at room 
temperatures.  

In the present system, where the control of protein's motion is realized via
dynamical charging of nanotubes, we can also consider these two generic 
(hinge and shear) configurations, shown schematically in Fig.~\ref{BIONA2}. 
Since, the tubes are different and become charged in equilibrium, the 
proteins should be able to {\it distinguish} them and deposit on 
them {\it asymmetrically} (see Figs.~\ref{BIONA1}-\ref{BIONA2}). In the bend 
configuration (left), the trajectory of the controlled protein domain is 
practically vertical, toward one of the tube's centers. In the shear 
configuration (right), the trajectory of the domain goes approximately 
in parallel with the vector connecting the tube's centers of masses. 

We assume that the balance of internal forces in the protein, in the presence 
of equilibrium charging of the tubes, adjusts the charged tip to a 
position ${\bf r}_0=(x_0,y_0)$. The dependence of the protein energy 
around (close to) this position can be considered to be parabolic
\begin{eqnarray}
V_R({\bf r}) \approx C_x\, (x-x_0)^2+ C_y\, (y-y_0)^2\, ,
\label{VR}
\end{eqnarray}
where the constants $C_x$, $C_y$ describe rigidity of the deformed protein 
\cite{Field}. Their values should be such, that the difference in 
energies $\Delta_E$ between the used conformations is kT$_{B}< \Delta_E< 
100$ kJ/mol (1 eV), where the last value is the lower energy limit required 
to deform {\it individual} protein domains \cite{Field}. 

\section{Discussion of the protein motion}

In order to estimate which of the configurations, in Fig.~\ref{BIONA2},
can be more easily controlled, we calculate the {\it distance} over which the 
protein domains move during the NIR-radiation induced charge transfer. 
The tip moves from the equilibrium position ${\bf r}_0$ to a new position 
${\bf r}_0^{'}$, given by the local minimum of the total potential
\begin{eqnarray}
V_T({\bf r})=V_C({\bf r})+V_S({\bf r})+V_R({\bf r})\, .
\label{VT}
\end{eqnarray}

\begin{figure}
\vspace{-3mm}
\hspace*{0mm}
{\hbox{\epsfxsize=90mm \epsffile{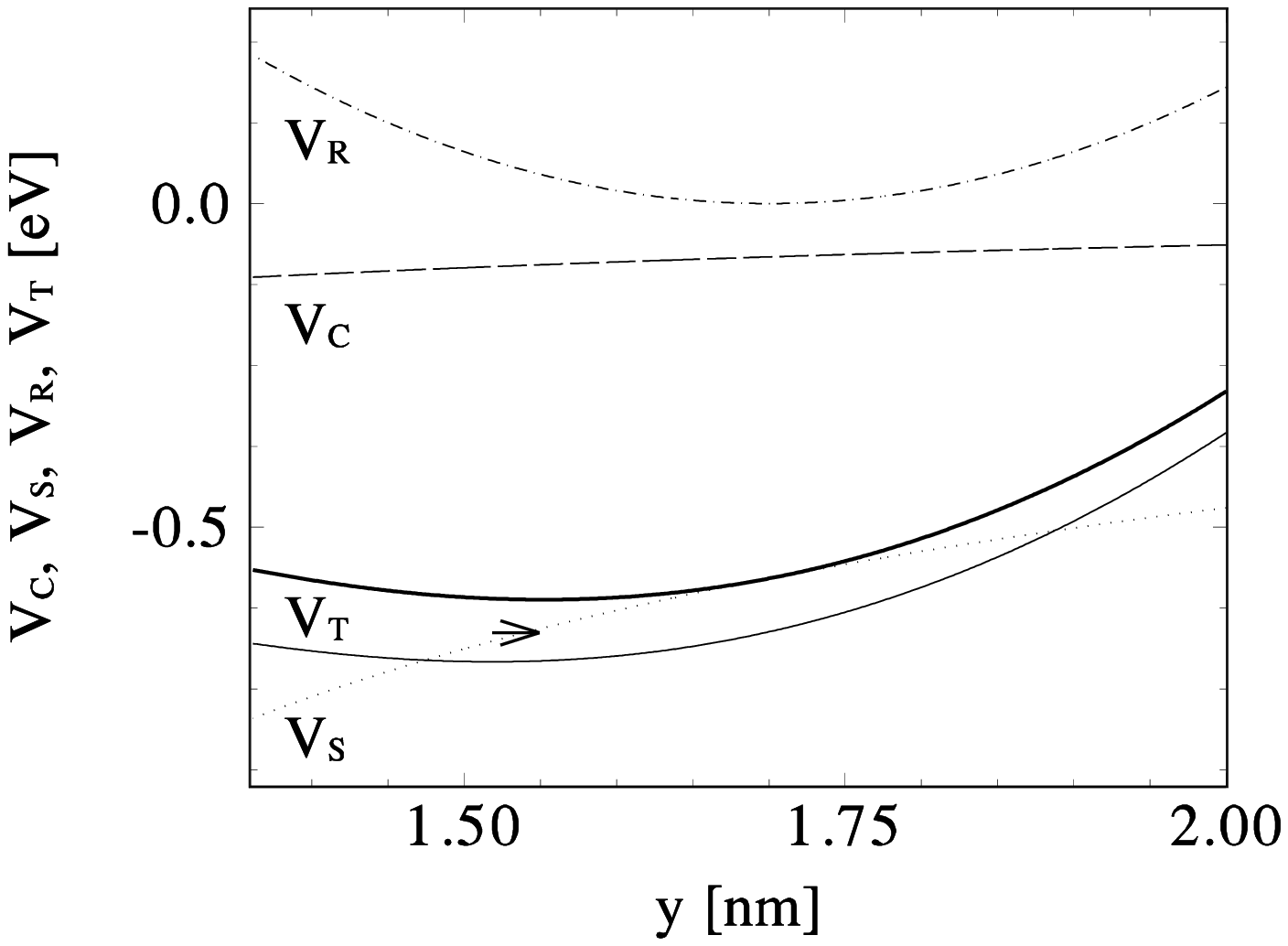} }}\\
\vspace{-8mm}
\hspace*{0mm}
{\hbox{\epsfxsize=90mm \epsffile{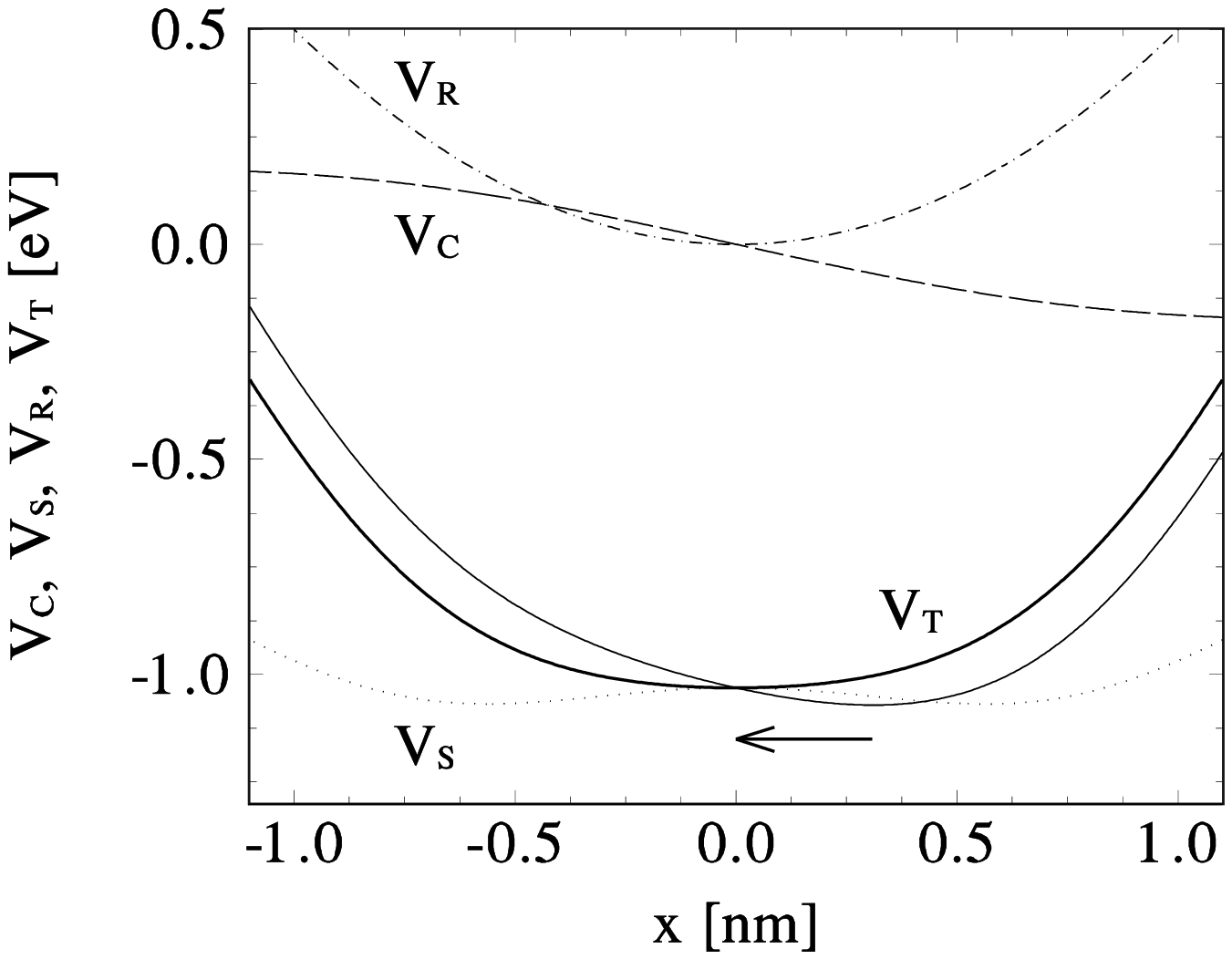} }}
\vspace{-10mm} 
\caption{The calculated potentials of the proteins in the hinge (up) 
and shear (down) configurations for the parameters in the text. From the 
local minima of $V_T$, corresponding to the change $\Delta \sigma =-0.1$ 
e/nm of the charge density on the tubes, we obtain the protein displacement 
$\approx 0.04$ nm and $\approx 0.35$ nm, respectively, shown by the arrows. }
\label{BIONA3}
\end{figure}

In Fig.~\ref{BIONA3} (up), we search this minimum for the ``hinge" 
configuration, shown in Fig.~\ref{BIONA2} (left). The tube centers are 
located at ${\bf r}_{1,2} =(x_{1,2},y_{1,2})$, $x_{1,2}= \mp 1.1\, r_T$, 
$y_{1,2}=0$.  We present the dependence of the potentials $V_C$, $V_S$, 
$V_R$ and $V_T$ on the $y$ distance from the center of the right nanotube, 
and  assume that a unit charge $q=e$ is at the tip of the domain \cite{Benson}. 
The results are calculated for the (effective) charge density $\sigma=0.1$ 
e/nm (equilibrium) and $\sigma=0$ (irradiation). We position the 
domain tip at $x_0=x_2$, $y_0=2.5\, r_T$ and use the rigidity constant 
$C_y=2$ eV/nm$^{-2}$. We can see that the {\it change} of the $V_C$ potential 
energy, due to the induced charge transfer, is small, while $V_S$ is rather 
large. With the above parameters, the $V_R$ potential can {\it locally} 
compensate the steep $V_S$, so that close to the tubes their sum is almost
flat. Then, the weak $V_C$ potential can control the position of the local 
minimum in $V_T$, but the overall motion is quite small (thin and thick 
solid line correspond to equilibrium and irradiation, respectively). The
magnitude of motion could be enlarged, if we let the $V_T$ potential 
to loose the local minimum in the {\it absence} of irradiation. Then the 
domain position would {\it fluctuate} from being adjacent to the tube 
to being almost at ${\bf r}_0$.

Such a large motion can be obtained directly in the ``shear" 
configuration, shown in Fig.~\ref{BIONA2} (right).  Here, the in-plane 
(of the tubes) components of the screening forces largely 
cancel each other, and the out-of-plane components are not effective. Thus 
the system responds more sensitively to the charging given by $V_C$, as we
show in Fig.~\ref{BIONA3} (down). We use parameters, $x_0=0$, $y_0=1.7\, 
r_T$ and $C_x=0.5$ eV/nm$^{-2}$, so that $V_R$ can {\it flatten} 
the two-well minima 
of $V_S$. In equilibrium, the tube charging causes that $V_T$ develops
a minimum, close to one of the $V_S$ minima.  During the irradiation, the 
charging decreases and the minimum shifts to $x=0$.  Since the two
conformations are shifted in energy by $\Delta_E \approx 50$ meV, they 
would not flip one to another at room temperatures kT$_{B} <30$ meV. This
domain motion is large enough to control the protein (enzyme) activity. 
It could open or block pockets on the ``back side" of the protein, that 
is not exposed to the nanotube, and change the catalytic strength of the 
proteins.  

We have demonstrated that NIR-radiation excited hybrid nanotube ropes could 
control the activity of proteins {\it in vivo}. Nanotube systems might 
also {\it directly} activate chemical reactions used in phototherapy 
\cite{MacDonald}, in particular, if special ``porphyrin-like" defects are
formed in the nanotube walls.  {\it In vitro}, one could also bias 
nanotubes externally in order to control biochemical reactions or use 
them in other applications on the nanoscale \cite{Baughman}. 

\vspace{3mm}

\noindent
{\bf Acknowledgments}
\vspace{-7mm}

\noindent
The author would like to thank S. Gross and E. J. Mele for valuable 
discussions. EU COCOMO is acknowledged for a support.


\end{document}